\documentclass[twocolumn,aps,prl,showpacs,raggedbottom,nobalancelastpage,amssymb,superscriptaddress]{revtex4-1}

\usepackage{graphicx}
\usepackage{amsmath}
\usepackage{amssymb}
\usepackage{bm}
\usepackage{color}
\usepackage{amsfonts}
\usepackage{appendix}
\usepackage{color}
\usepackage{dsfont}
\usepackage{soul}
\setstcolor{red}

\def\re{\mathrm{Re}}
\def\im{\mathrm{Im}}

\definecolor{lightgrey}{rgb}{0.7,0.7,0.7}

\newcommand{\up}{\uparrow}
\newcommand{\dn}{\downarrow}

\newcommand{\bra}[1]{\left<{#1}\right|}
\newcommand{\ket}[1]{\left|{#1}\right>}
\newcommand{\braket}[2]{\left<\left.{#1}\right|{#2}\right>}

\newcommand{\ignore}[1]{\relax}

\begin{document}
\title{
Charge sensing amplification via weak values measurement
}
\author{Oded Zilberberg}
\affiliation{Department of Condensed Matter Physics, The Weizmann Institute of Science, Rehovot 76100, Israel.}
\author{Alessandro Romito}
\affiliation{Institut f\"{u}r Theoretische Festk\"{o}rperphysik, Universit\"{a}t Karlsruhe, D--76128 Karlsruhe, Germany.}
\affiliation{Dahlem Center for Complex Quantum Systems and Fachbereich Physik, Freie Universit\"at Berlin, 14195 Berlin, Germany}
\author{Yuval Gefen}
\affiliation{Department of Condensed Matter Physics, The Weizmann Institute of Science,  Rehovot 76100, Israel.}

\received{23 September 2010}
\published{February 24, 2011}
\begin{abstract}
A protocol employing weak values (WVs) to obtain ultra sensitive amplification of weak signals in the context of a solid-state setup is proposed. We consider an Aharonov-Bohm interferometer where both the orbital and the spin degrees of freedom are weakly affected by the presence of an external charge to be detected. The interplay between the spin and the orbital WVs leads to a significant amplification even in the presence of finite temperature, voltage, and external noise.
\end{abstract}


\maketitle
Weak values (WVs) were introduced more than 20 years ago~\cite{Aharonov:1988aa} as a peculiarity of quantum mechanics.
A WV measurement consists in
(i) initializing the \emph{system} in a certain state $\ket{\psi_i}$---\emph{preselection},
(ii) coupling weakly an observable $\hat{A}$ of the system with an observable $\hat{B}$ of the \textit{detector} (via a von Neumann interaction~\cite{von-neumann}),
and (iii) retaining the detector output only if the system is eventually measured to be in a chosen final state, $\ket{\psi_f}$---\emph{postselection}.
The average signal detected by the detector will then be proportional to the real or imaginary part of the complex WV,
${}_f\langle \hat{A}\rangle_i=\bra{\psi_f}\hat{A}\ket{\psi_i}/\braket{\psi_f}{\psi_i}$,
rather than to the standard average value, $\bra{\psi_i} \hat{A} \ket{\psi_i}$. Further discussion of the context in which WV should be understood has been provided~\cite{Wiseman:2002,Jozsa:2007,Dressel:2010}.

Going beyond the peculiarities of WV protocols,  recent series of works explored the   potential of WVs in quantum optics~\cite{Ritchie:1990,Pryde:2005,Hosten:2008,Dixon:2009,Starling:2009,Brunner:2010,Starling:2010b} and solid-state physics~\cite{Williams:2008,Romito:2008,Shpitalnik:2008}, ranging from experimental observation to their application to hypersensitive measurements.
In the latter, a measurement, performed by a detector \emph{entangled} with a system whose states can be preselected and postselected, leads to an amplified signal in the detector that enables sensing of small, otherwise inaccessible quantities, e.g. sensing the deflection angle of a mirror of the order of $\sim 500$ frad~\cite{Dixon:2009}.
Within such a WV amplification protocol only a subset of the detector's readings, associated with the tail of the signal's
distribution, is accounted for. Yet, the large value of ${}_f\langle \hat{A}\rangle_i$ outweighs the scarcity of the data points
and leads to signal-to-noise ratio (SNR) amplification~\cite{Starling:2009}.

\begin{figure}[ht!]
\includegraphics[width=7.5cm]{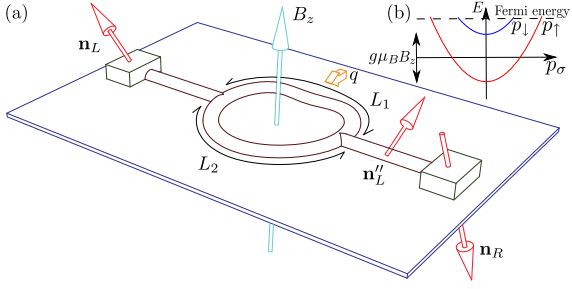}
\caption[]{\label{Fig:1}
(Color online) (a) A sketch of the WV hyper-sensitive charge measurement setup. It consists of a half-metal open AB interferometer with Zeeman magnetic field ${\bf B}=B_z{\bf e}_z$. The half-metal left and right leads with spin orientations $\hat{\mathbf{n}}_L,\hat{\mathbf{n}}_R$, respectively, serve as a SV that measures the spin orientations of the electrons. The length of the upper (lower) arm is $L_1\; (L_2)$. The spin orientation $\hat{\mathbf{n}}_L''$ exits the right junction of the interferometer. A small charge, $q$, is situated next to one of the arms of the interferometer, weakly changing the confining geometry of this arm. Consequently, the electron trajectory and momentum of electrons passing through this arm are modified, inducing additional electron spin rotation which depends on whether the
upper or the lower interferometer's arm is traversed (i.e., the spin is coupled to the {}``which-path coordinate''). The signal due
to $q$  is stored in the spin state which is read in the current through the SV. (b) A sketch of the dispersion curve for $\hat{\cal H}$, cf.  Eq.~\eqref{eq:H}.
}
\end{figure}

Here we present a paradigm WV hyper-sensitive measurement in the context of solid state systems.
It consists of an open semiconducting Aharonov-Bohm (AB) interferometer subject to a Zeeman magnetic field contacted to half-metallic (strong ferromagnetic) leads. Such a device is employed to sense a small charge, $q$, situated next to one of the arms of the interferometer [see Fig.~\ref{Fig:1}(a)];
$q$ affects the electron trajectory and momentum in this interferometer arm. While spinless AB interferometers have been fully characterized as detectors (e.g., Ref.~\cite{Schomerus:2007}), here we harness the additional (spin) degree of freedom (d.o.f.) for amplified detection. Within our conceptual scheme, $q$ may be thought of as the uncompensated charge induced by a gated electrode. As such, $q$ will be treated classically and induce an electron spin rotation due to the altered orbital motion of the electron (mind the magnetic field).
We thus have two d.o.f. --- orbital and spin --- which serve as {}``amplifier'' and {}``detector'' respectively (or vice versa). Our amplification scheme, involving these d.o.f., is compared with a simple-minded scheme where (in the absence of interferometry) only the spin d.o.f. is involved.

 The value of $q$ (its weak effect on the interferometer)  is read in the current through the half-metal drain acting as a spin valve (SV).
We show that a properly chosen preselection and postselection of interferometer states, while reducing the current at the drain, makes the spin coordinate of the transmitted electrons hypersensitive to the small charge. Our analysis underlines the interplay between spin-related and orbital-related WVs. We show that even when the orbital WV, marking the amplification of the current signal absorbed in the interferometer's drain, is countered by the reduced current, our protocol can still be utilized to enhance signal-to-external-noise ratio. Our protocol may be extended to realistic multiterminal setups that can be employed experimentally.

We begin by describing the transport through a wire connected to SV leads. The electrons' motion is ballistic,
described by the Hamiltonian
\begin{align}
\hat{\cal H} =
\frac{1}{2m} (\hat{\bf{p}}-\frac{e}{c}{\bf A})^2
+ \frac{g\mu_B}{2} {\bf B}\cdot\hat{\bf \sigma}\,.
\label{eq:H}
\end{align}
Here ${\bf A}$ represents an AB  vector potential and ${\bf B}=B_z{\bf e}_z$ is an {\it additional} magnetic field~\cite{footnote:general-magnetic}.
In generalized cylindrical coordinates $[\tilde{r}(\theta),\theta,z]$~\cite{Whitney:2008,Meijer:2002}, the Hamiltonian reads
$\hat{\cal H}_{\rm cyl}=(1/2m)\hat p_\theta^2 + E_0+(g\mu_B/2)B_z \hat \sigma_z$,
where $E_0$ is the lowest transverse mode's energy.
The eigenmodes' momenta along the wire, $p_\sigma\equiv p_\theta(\sigma)$, are given by
$p_\sigma=  \pm\sqrt{ p_0^2 - g\mu_B m B_z \sigma  }$,
where $p_0 = [2m(E_{\rm F}-E_0)]^{1/2}$, and $\sigma=\pm 1=(\up,\dn)$ labels the spin eigenstates, $\ket{\uparrow}$, $\ket{\downarrow}$, in the direction of the applied magnetic field [see Fig.~\ref{Fig:1}(b)].

The current, $I$, through the ballistic device is given by the Landauer-B\"uttiker formula~\cite{Landauer:1957}
\begin{align}
I&=\frac{e}{h}\int dE \, \left|t_{L\rightarrow R} (E) \right|^2 [f_L(E)-f_R(E)] \,,
\label{landauer}
\end{align}
where $t_{L\rightarrow R}(E)$ is the transmission amplitude through the device at energy $E$ and $f_{L(R)}(E)$ is the Fermi distribution functions of the left (right) lead.
Let us first discuss the effect of $q$ in the case of an energy independent transmission $t_{L\rightarrow R}(E) = t_{L \rightarrow R} \equiv t_{L \rightarrow R}(E_F)$.
An electron injected in a wire
with spin $\sigma= \pm$, traversing a length L, will acquire a phase $\vartheta_{\sigma} = p_{\sigma} L/\hbar$.
As a result, an electron injected at energy $E_F$ with spin $\ket{\mathbf{n}_L}$ precesses in the magnetic field to a new spin orientation $\ket{\mathbf{n}_L^\prime}=U(L)\ket{\mathbf{n}_L}=e^{i (\Delta p L/\hbar) \hat{\sigma}_z}\ket{\mathbf{n}_L}$, where we have introduced $2 \bar{p} \equiv p_{\uparrow} + p_{\downarrow}$ and $2 \Delta p \equiv p_{\uparrow} - p_{\downarrow}$.
The ferromagnetic leads act as a SV, detecting the final spin orientation, with the transmission $T=|t_{L \rightarrow R}|^2 =  |\langle \mathbf{n}_R|\mathbf{n}_L'\rangle|^2$, where $\ket{\mathbf{n}_R}$ is the spin orientation of the right lead.

The effect of the charge $q$ situated in the vicinity of the wire can be deduced from a semiclassical analysis~\cite{supplementary}.
We find that to first order in $q$, the presence of the charge induces an additional spin precession
\begin{align}
 U_{\text{int}}\ket{\mathbf{n}_L^\prime}\approx e^{iq\eta\Delta p\hat{\sigma}_z}\ket{\mathbf{n}_L^\prime}\approx  \left(1+i q \eta\Delta p\hat{\sigma}_z\right)\ket{\mathbf{n}_L'}\, ,
 \label{sv_spin_effect}
 \end{align}
  where $\eta$ encodes the details of the setup~\cite{supplementary}.
The current in the drain is sensitive to this extra spin rotation induced by $q$. We define the {}``signal'' of $q$ as $\Delta T_q =  T_{q}- T_{q=0}$. For this simple SV case
\begin{align}
\Delta T_q=  2 q\eta\Delta p  |\langle \mathbf{n}_R | \mathbf{n}_L' \rangle |^2 \im\{ {}_R\langle \hat{\sigma}_z\rangle_L\}\, ,
\label{sv_signal}
\end{align}
 where we define the spin WV:
\begin{align}
    {}_R\langle \hat{\sigma}_z\rangle_L&=\frac{\bra{\mathbf{n}_R}\hat{\sigma}_z\ket{\mathbf{n}_L^\prime}}{\braket{\mathbf{n}_R}{\mathbf{n}_L^\prime}}=\frac{\sum_\sigma \sigma \braket{\mathbf{n}_R}{\sigma}\braket{\sigma}{\mathbf{n}_L^\prime}}{\sum_\sigma \braket{\mathbf{n}_R}{\sigma}\braket{\sigma}{\mathbf{n}_L^\prime}}\, .
    \label{spinWV}
\end{align}
Note that the signal $\Delta T_q$ is greatly reduced with the vanishing of the $|\langle \mathbf{n}_R | \mathbf{n}_L' \rangle |^2$ factor.
The sensitivity of the measurement is obtained by comparing the signal with its uncertainty due to an extraneous noise source. As an example  we assume an uncertainty in $\mathbf{n}_R $, i.e. $|\mathbf{n}_R\rangle =\exp[i\xi \mathbf{n}_\xi \cdot\mathbf{\sigma}]|\mathbf{n}_R\rangle $, where $\xi$ fluctuates much slower than the time of flight of electrons in the device, and $\langle \xi \xi \rangle =\Delta_\xi^2$.
This leads to an error in the transmission $\Delta T_{\xi} = 2 |\langle \mathbf{n}_R | \mathbf{n}_L' \rangle|^2 \Delta_\xi  \im\{ {}_R \langle \sigma_{\xi}\rangle_{L} \}$. The SNR is therefore
\begin{equation}
    \label{segnale-rumore}
    \alpha_{\text{SV}} \equiv \frac{\left|\Delta T_{q}\right|}{\left|\Delta T_{\xi}\right|}=\left|q  \eta \frac{\Delta p} {\Delta_{\xi}} \frac{\im\{ {}_R\langle \hat{\sigma}_z\rangle_L\}}{\im\{ {}_R \langle \sigma_{\xi}\rangle_{L} \}}\right| \, ,
\end{equation}
which exhibits no amplification.

We now combine the spin d.o.f. with an interferometer geometry (see Fig.~\ref{Fig:1}). It is equivalent to including an additional orbital d.o.f. which is affected by $q$. Our interferometer is sufficiently open such that no higher windings around it occur (or multiterminal).
Passing through the interferometer's arms, $L_1$, $L_2$, an electron with spin $\sigma$ has transmission  amplitudes $t_{1\sigma} = |t_1| e^{i(p_\sigma L_1/\hbar+\varphi_1)}$, $t_{2\sigma} = |t_2| e^{i(p_\sigma L_2/\hbar+\varphi_2)}$, respectively.
Hence, the transmission through our device can be written as a spin scalar product
$T=|t_{L \rightarrow R}|^2 = \mathcal{N}^2|\langle \mathbf{n}_R | \mathbf{n}_L'' \rangle |^2$,
where $\ket{\mathbf{n}_L''} = \langle \phi_f | U_{\text{int}} | \phi_i \rangle \ket{\mathbf{n}_L'}/\mathcal{N}$ is a properly normalized spin state that exits the right junction  of the interferometer with $\mathcal{N} = \sqrt{\left|\bra{\phi_f}  U_{\text{int}} \ket{\phi_i} \ket{\mathbf{n_L}'}\right|^2} $.
Here, we have included the purely orbital effect of the interferometer (amplifier) by defining a state that enters into the right junction of the interferometer (preselection), $\ket{\phi_i} = |t_1| e^{i(\bar{p} L_1/\hbar+\varphi_1)}  \ket{1} + |t_2| e^{i(\bar{p} L_2/\hbar+\varphi_2)}  \ket{2}$, and a state that comes out of it (postselection), $\ket{\phi_f} = e^{i \Phi_{\text{AB}}}\ket{1} + \ket{2}$, where $\Phi_{\rm{AB}}=-eB_z{\cal A}/(g \mu_{\rm B}\hbar)$ and $\cal A$ is the enclosed area in the AB ring. The state $\ket{1}$ ($\ket{2}$) denotes an orbital wavefunction at the origin of arm $1$ ($2$).
The  spin rotation is the result of two contributions: First, provided the two interferometer arms are of equal length, $L_2$, the precession in the applied magnetic field yields   $\ket{\mathbf{n}_L'}=U(L_2)\ket{\mathbf{n}_L}$. Second, there is an extra rotation of the component that runs through arm $L_1$, given by (an interplay of spin and orbit) $U_{\text{int}} = e^{i (\Delta p \Delta L/\hbar)\hat{A} \hat{\sigma}_z}$, with $\Delta L=L_1-L_2$ and $\hat{A}=\ket{1}\bra{1}$. Henceforth, we refer to $\hat{A}$ as the which-path operator. The emerging rotated spinor is $| \mathbf{n}_L'' \rangle$.

The effect of $q$ situated in the vicinity of arm $1$ can be written in an operator form~\cite{supplementary}
\begin{equation}
    \label{carica}
    U_{\text{int}} \rightarrow  U_{\text{int}}^q = e^{-i q \hat{A}\left[(\delta+\eta\bar{p})-\eta\Delta p\hat{\sigma}_z\right]} U_{\text{int}}^{q=0} \, ,
\end{equation}
where we included the effect of the charge on the enclosed area, hence the enclosed flux $\Phi_{\text{AB}}\rightarrow \Phi_{\text{AB}}+\delta e$.
For simplicity, we explicitly discuss the results in the case $(\Delta p \Delta L/\hbar) =0 \mod(2 \pi)$, in which the operator $U_{\text{int}}^{q=0}=\mathds{1}$ and the amplifier and detector are coupled only due to the presence of $q$~\cite{footnote:coupled}.
In a weak measurement regime the response of the detector is linear in $q$. Expanding the exponent in Eq.~\eqref{carica}, the change in the transmission to linear order in $q$ is
\begin{eqnarray}
    \label{trasmissione}
    & &\Delta T_q = -
    2 q |\langle \mathbf{n}_R | \mathbf{n}_L' \rangle |^2 |\langle \phi_f |\phi_i \rangle|^2  \nonumber \\
    &  & \times  [ (\delta+\eta\bar{p}) \im\{ {}_f \langle \hat{A}\rangle_i \} -\eta\Delta p \im\{ {}_f\langle \hat{A}\rangle_i {}_R\langle \hat{\sigma}_z\rangle_L\}]\, ,
    \end{eqnarray}
where we have introduced the orbital WV
\begin{align}
\label{WPWV}
    {}_f\langle \hat{A}\rangle_i &=\frac{\bra{\phi_f}\hat{A}\ket{\phi_i}}{\braket{\phi_f}{\phi_i}}=\frac{t_1}{t_1+t_2 e^{i\tilde{\Phi}}} \, ,
\end{align}
with $\tilde{\Phi} = \Phi_{\rm{AB}}- (\bar{p} \Delta L/\hbar) + \varphi_2 - \varphi_1$.

In order to appreciate the enhanced sensitivity due to the post-selection in the interferometer, we focus on the simplest case ${}_f \langle \hat{A}\rangle_i \in \mathbb{R}$, where only the second term in Eq.~(\ref{trasmissione}) is present:
\begin{eqnarray}
    \label{trasmissionereal}
    \Delta T_q =
    2 q\eta\Delta p  |\langle \mathbf{n}_R | \mathbf{n}_L' \rangle |^2 |\langle \phi_f |\phi_i \rangle|^2
     {}_f\langle \hat{A}\rangle_i\im\{{}_R\langle \hat{\sigma}_z\rangle_L\}\, .
\end{eqnarray}
In this case the  spin emerging at the  right end of the interferometer  is
\begin{align}
	\ket{\mathbf{n}_L''} \approx  \frac{\langle\phi_f |\phi_i \rangle}{|\langle \phi_f |\phi_i \rangle|}\left(1+i q\eta\Delta p {}_f \langle \hat{A}\rangle_i\hat{\sigma}_z\right)\ket{\mathbf{n}_L'}\, .
\end{align}
Comparing this with Eq.~\eqref{sv_spin_effect}, we see that the spin change due to the nearby charge is amplified by the WV factor ${}_f \langle \hat{A}\rangle_i$. Thus, spin rotation due to the presence of $q$ is amplified by the WV procedure.

Here, however, rather than measuring directly the spin rotation, we measure the signal, $\Delta T_q$. The latter is affected by both the orbital and spin d.o.f..
Comparing Eq.~\eqref{trasmissionereal} with Eq.~\eqref{sv_signal}, we see that the amplification due to the orbital WV in  Eq.~\eqref{WPWV} is compensated by a reduction prefactor, $|\braket{\phi_f}{\phi_i}|^2$.
Tuning the interferometer to be destructive on the right junction, leading to a large ${}_f\langle \hat{A}\rangle_i$, will be countered by the reduced current.

\begin{figure}[ht!]
\includegraphics[width=7.5cm]{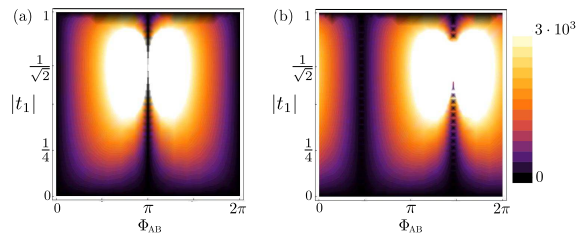}
\caption[]{\label{Fig:2}
(Color online) Density plot of the amplification factor, $\alpha_{\text{SV+interf}}/\alpha_{\text{SV}}$, in the case of ${}_f\langle \hat{A}\rangle_i \in \mathbb{C}$ for (a) $(\Delta p \Delta L/\hbar) =0 \mod(2 \pi)$, and (b) $(\Delta p \Delta L/\hbar) = 2.6\cdot 10^{-3}$. In both plots $\mathbf{n}_L=\mathbf{n}_R=\sigma_x$, $B_z=1$T, $(\bar{p}/\hbar) \sim 7\cdot 10^{-2}$(1/nm), $(\Delta p/\hbar) \sim 1\cdot 10^{-4}$(1/nm), $\hbar e \eta \sim 20.7$nm,  $\delta e \sim 1.5\cdot 10^{-2}$~\cite{supplementary}.}
\end{figure}

The \emph{relative} effect of the charge $q$, i.e. $(T_{q}-T_{q=0})/T_{q=0}$, is nevertheless enhanced by a large prefactor ${}_f\langle \hat{A}\rangle_i$. In order to appreciate the added sensitivity due to the orbital WV, we consider the response of this device to the same external noise as before
$ \Delta T_{\xi} = 2 \Delta_\xi |\langle \mathbf{n}_R | \mathbf{n}_L' \rangle|^2 |\langle \phi_f |\phi_i \rangle|^2  \im\{ {}_R \langle \sigma_{\xi}\rangle_{L} \}$. Assuming, again, ${}_f \langle \hat{A}\rangle_i \in \mathbb{R}$, we obtain a SNR
\begin{equation}
    \label{amplificazione}
    \alpha_{\text{SV+interf.}} = \left|{}_f\langle \hat{A}\rangle_i\right| \alpha_{\text{SV}} \, .
\end{equation}
Since the WV, ${}_f\langle \hat{A}\rangle_i$, can be arbitrarily large, it is possible to amplify the SNR at will.
Indeed the postselection due to the interferometer reduces the final current to be measured, but reduces even more the relative uncertainty on the current due to the noise-induced-error in the SV orientation [see Fig.~\ref{Fig:2}(a)].
The price to be paid for the amplification is that one has to detect smaller currents, and this sets a technical bound for the amplification.

Beyond the case ${}_f \langle \hat{A}\rangle_i \in \mathbb{R}$, the current through the device [cf. Eq.~\eqref{trasmissione}] has two terms.
The second term contains the interplay between spin and interferometer d.o.f. leading to the WV amplification we discussed, while the first term is equivalent to the effect of the charge on a spinless AB interferometer times a reduction prefactor due to the spin.
The characterization of the amplification effect in terms of the SNR, as in Eqs.~(\ref{segnale-rumore},\ref{amplificazione}), is valid in the general case [${}_f\langle \hat{A}\rangle_i \in \mathbb{C}$, $(\Delta p \Delta L/\hbar) \neq 0 \mod( 2 \pi)$]. We depict the amplification factor for such a general case in Fig.~\ref{Fig:2}(b).

\begin{figure}[ht!]
\includegraphics[width=7.5cm]{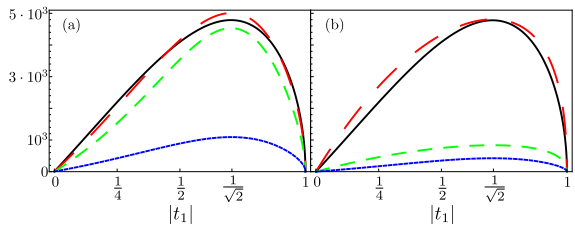}
\caption[]{\label{Fig:3}
(Color online) (a) The amplification factor, $\alpha_{\text{SV+interf}}/\alpha_{\text{SV}}$, as a function of the transmission through arm $1$ for different values of the temperature (a) and voltage bias  (b).
In (a), $V=0.1$meV, and $T=10$mK (solid black curve),  $T=25$K (long-dashed red curve), $T=50$K (medium-dashed green curve), $T=100$K (short-dashed blue curve).
In (b), $T=10$mK, and  $V=0.1$meV (solid black curve), $V=10$meV (long-dashed red curve), $V=25$meV (medium-dashed green curve), $V=50$meV (short-dashed blue curve).
All plots $\Phi_{\text{AB}}=\pi$ and all other parameters are chosen as in Fig.~\ref{Fig:2}.
}
\end{figure}

At finite temperature we cannot neglect the energy dependence of the transmission amplitude, $t_{L\rightarrow R}\rightarrow t_{L\rightarrow R}(E)$, and
one needs to perform the integral over energy in Eq.~(\ref{landauer}). We linearize the electron energy spectrum around the Fermi energy. The energy-dependent transmission is determined by Eq.~\eqref{trasmissione} where $\bar{p}\rightarrow \bar{p}(E)=\bar{p}[1+(m/\mathcal{P}^2)E]$, $\Delta p\rightarrow \Delta p(E)=\Delta p[1-(m/\mathcal{P}^2)E]$~\cite{supplementary}.
The WV amplification in ($\alpha_{\text{SV+interf}}/\alpha_{\text{SV}}$) is gradually suppressed at higher temperatures and voltage bias ($k_B T \sim eV \gtrsim (\mathcal{P}^2/m)\max\{ (\hbar/\Delta p L), (\hbar/\bar{p} \Delta L) \}$), see Fig.~\ref{Fig:3}.
At $ k_B T \lesssim (\mathcal{P}^2/m)\max\{ (\hbar/\Delta p L), (\hbar/\bar{p} \Delta L) \}$ the SNR can even be enhanced since temperature affects the signal and the noise differently.

\begin{figure}[ht!]
\includegraphics[width=7.5cm]{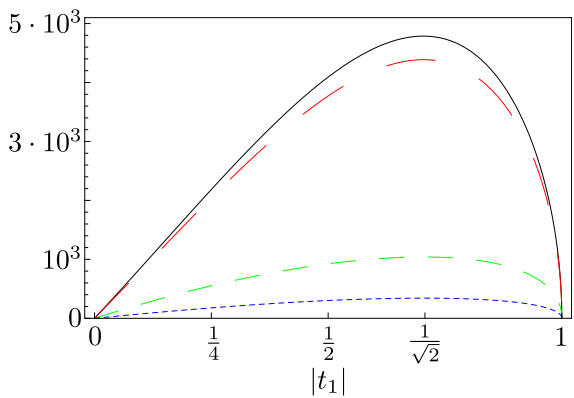}
\caption[]{\label{Fig:4}
(Color online) The amplification factor, $\alpha_{\text{SV+interf}}/\alpha_{\text{SV}}$, as a function of the transmission through arm $1$ for different strength of the magnetic field fluctuations: $\Delta B=10$mT (solid black curve), $\Delta B=100$mT (long-dashed red curve), $\Delta B=500$mT (medium-dashed green curve), and $\Delta B=700$mT (short-dashed blue curve). All plots are for $\Phi_{\text{AB}}=\pi$ and all the other parameters are chosen as in Fig.~\ref{Fig:2}.}
\end{figure}

In much the same way,  Gaussian magnetic field fluctuations of width $\Delta B$ will smear the WV amplification. We include the dependence of the transmission amplitude, $t_{L\rightarrow R}\rightarrow t_{L\rightarrow R}(B)$, and
integrate over the magnetic field distribution. In a way compatible with our earlier approximation  we linearize the electron energy spectrum around the tuned magnetic field $B_0$, in which case the magnetic field-dependent signal is determined by Eq.~\eqref{trasmissione}, where $\bar{p}\rightarrow \bar{p}(B)=\bar{p}-\Delta p(g\mu_B m/2\mathcal{P}^2)B$, $\Delta p\rightarrow \Delta p(B)=\Delta p+\bar{p}(g\mu_B m/2\mathcal{P}^2)B$~\cite{supplementary}.
As shown in Fig.~\ref{Fig:4}, the amplification due to the WV is completely suppressed in the high magnetic noise limit $\Delta B  \Phi_{\textrm{AB}} \gg B$. Unlike the temperature or voltage case, here the WV amplification is suppressed predominantly by the effect of fluctuations on the AB phase, rather than dephasing due to the momenta, $\Delta p$, $\bar{p}$.

Here we have proposed a setup which exploits the notion of ultrahigh amplification using a weak value protocol, in the context of a quantum solid-state device.
An important feature of our design is that by assigning to spin and orbital d.o.f. the meaning of a detector and amplifier that undergoes preselection and postselection, it allows us to observe WVs without
synchronizing (in time) pulses in the two {}``devices''. Such a necessity arose  in earlier proposals~\cite{Williams:2008,Romito:2008,Shpitalnik:2008}. We stress that the main focus of this work was to demonstrate conceptually that WV amplification is possible in solid-state devices. While an attempt has been made to conform to realistic values of parameters~\cite{supplementary}, the present analysis does not purport to substitute a careful numerical or material science oriented analysis of an operating device. Some of the ingredients that need to be accounted for are the inclusion of multichannel wires, and the fact that in practice the readout of the device will need to be calibrated against known values of $q$ (or values of a charge producing gate voltage)~\cite{Hosten:2008,Dixon:2009}.

This work was supported by GIF, DFG Center for
Functional Nanostructures, Einstein Minerva Center,
US-Israel BSF, ISF, Minerva Foundation, Israel-Korea MOST grant, EU GEOMDISS, Israel MOIA grant, and the Alexander von Humboldt Foundation. We would like to thank B. Nissan-Cohen and D. Klarman for fruitful discussions and R. Eliav for the illustration of the device.


\begin{thebibliography}{23}
\expandafter\ifx\csname natexlab\endcsname\relax\def\natexlab#1{#1}\fi
\expandafter\ifx\csname bibnamefont\endcsname\relax
  \def\bibnamefont#1{#1}\fi
\expandafter\ifx\csname bibfnamefont\endcsname\relax
  \def\bibfnamefont#1{#1}\fi
\expandafter\ifx\csname citenamefont\endcsname\relax
  \def\citenamefont#1{#1}\fi
\expandafter\ifx\csname url\endcsname\relax
  \def\url#1{\texttt{#1}}\fi
\expandafter\ifx\csname urlprefix\endcsname\relax\def\urlprefix{URL }\fi
\providecommand{\bibinfo}[2]{#2}
\providecommand{\eprint}[2][]{\url{#2}}

\bibitem[{\citenamefont{Aharonov et~al.}(1988)\citenamefont{Aharonov, Albert,
  and Vaidman}}]{Aharonov:1988aa}
\bibinfo{author}{\bibfnamefont{Y.}~\bibnamefont{Aharonov}},
  \bibinfo{author}{\bibfnamefont{D.~Z.}~\bibnamefont{Albert}}, \bibnamefont{and}
  \bibinfo{author}{\bibfnamefont{L.}~\bibnamefont{Vaidman}},
  \bibinfo{journal}{Phys. Rev. Lett.} \textbf{\bibinfo{volume}{60}},
  \bibinfo{pages}{1351} (\bibinfo{year}{1988}).

\bibitem[{\citenamefont{von Neumann}(1938)}]{von-neumann}
\bibinfo{author}{\bibfnamefont{J.}~\bibnamefont{von Neumann}},
  \emph{\bibinfo{title}{Mathematical Foundation of Quantum Theory}}
  (\bibinfo{publisher}{Princeton, University Press, New Jersey},
  \bibinfo{year}{1938}).

\bibitem[{\citenamefont{Wiseman}(2002)}]{Wiseman:2002}
\bibinfo{author}{\bibfnamefont{H.~M.}~\bibnamefont{Wiseman}},
  \bibinfo{journal}{Phys. Rev. A} \textbf{\bibinfo{volume}{65}},
  \bibinfo{pages}{032111} (\bibinfo{year}{2002}).

\bibitem[{\citenamefont{Jozsa}(2007)}]{Jozsa:2007}
\bibinfo{author}{\bibfnamefont{R.}~\bibnamefont{Jozsa}},
  \bibinfo{journal}{Phys. Rev. A} \textbf{\bibinfo{volume}{76}},
  \bibinfo{pages}{044103} (\bibinfo{year}{2007}).

\bibitem[{\citenamefont{Dressel et~al.}(2010)\citenamefont{Dressel, Agarwal,
  and Jordan}}]{Dressel:2010}
\bibinfo{author}{\bibfnamefont{J.}~\bibnamefont{Dressel}},
  \bibinfo{author}{\bibfnamefont{S.}~\bibnamefont{Agarwal}}, \bibnamefont{and}
  \bibinfo{author}{\bibfnamefont{A.~N.} \bibnamefont{Jordan}},
  \bibinfo{journal}{Phys. Rev. Lett.} \textbf{\bibinfo{volume}{104}},
  \bibinfo{pages}{240401} (\bibinfo{year}{2010}).

\bibitem[{\citenamefont{Ritchie et~al.}(1991)\citenamefont{Ritchie, Story, and
  Hulet}}]{Ritchie:1990}
\bibinfo{author}{\bibfnamefont{N.~W.~M.} \bibnamefont{Ritchie}},
  \bibinfo{author}{\bibfnamefont{J.~G.} \bibnamefont{Story}}, \bibnamefont{and}
  \bibinfo{author}{\bibfnamefont{R.~G.} \bibnamefont{Hulet}},
  \bibinfo{journal}{Phys. Rev. Lett.} \textbf{\bibinfo{volume}{66}},
  \bibinfo{pages}{1107} (\bibinfo{year}{1991}).

\bibitem[{\citenamefont{Pryde et~al.}(2005)\citenamefont{Pryde, O'Brien, White,
  Ralph, and Wiseman}}]{Pryde:2005}
\bibinfo{author}{\bibfnamefont{G.~J.}~\bibnamefont{Pryde}} \emph{et al.},
    \bibinfo{journal}{Phys. Rev. Lett.} \textbf{\bibinfo{volume}{94}},
  \bibinfo{pages}{220405} (\bibinfo{year}{2005}).
  
\bibitem[{\citenamefont{Hosten and Kwiat}(2008)}]{Hosten:2008}
\bibinfo{author}{\bibfnamefont{O.}~\bibnamefont{Hosten}} \bibnamefont{and}
  \bibinfo{author}{\bibfnamefont{P.}~\bibnamefont{Kwiat}},
  \bibinfo{journal}{Science} \textbf{\bibinfo{volume}{319}},
  \bibinfo{pages}{787} (\bibinfo{year}{2008}).

\bibitem[{\citenamefont{Dixon et~al.}(2009)\citenamefont{Dixon, Starling,
  Jordan, and Howell}}]{Dixon:2009}
\bibinfo{author}{\bibfnamefont{P.~B.} \bibnamefont{Dixon}} \emph{et al.}, \bibinfo{journal}{Phys. Rev. Lett.}
  \textbf{\bibinfo{volume}{102}}, \bibinfo{pages}{173601}
  (\bibinfo{year}{2009}).

\bibitem[{\citenamefont{Starling et~al.}(2009)\citenamefont{Starling, Dixon,
  Jordan, and Howell}}]{Starling:2009}
\bibinfo{author}{\bibfnamefont{D.~J.}~\bibnamefont{Starling}} \emph{et al.},
    \bibinfo{journal}{Phys. Rev. A} \textbf{\bibinfo{volume}{80}},
  \bibinfo{pages}{041803} (\bibinfo{year}{2009}).
  
\bibitem[{\citenamefont{Brunner and Simon}(2010)}]{Brunner:2010}
\bibinfo{author}{\bibfnamefont{N.}~\bibnamefont{Brunner}} \bibnamefont{and}
  \bibinfo{author}{\bibfnamefont{C.}~\bibnamefont{Simon}},
  \bibinfo{journal}{Phys. Rev. Lett.} \textbf{\bibinfo{volume}{105}},
  \bibinfo{pages}{010405} (\bibinfo{year}{2010}).

\bibitem[{\citenamefont{Starling et~al.}(2010)\citenamefont{Starling, Dixon,
  Williams, Jordan, and Howell}}]{Starling:2010b}
\bibinfo{author}{\bibfnamefont{D.~J.} \bibnamefont{Starling}} \emph{et al.}, \bibinfo{journal}{Phys. Rev. A}
  \textbf{\bibinfo{volume}{82}}, \bibinfo{pages}{011802}
  (\bibinfo{year}{2010}).

\bibitem[{\citenamefont{Williams and Jordan}(2008)}]{Williams:2008}
\bibinfo{author}{\bibfnamefont{N.~S.} \bibnamefont{Williams}} \bibnamefont{and}
  \bibinfo{author}{\bibfnamefont{A.~N.} \bibnamefont{Jordan}},
  \bibinfo{journal}{Phys. Rev. Lett.} \textbf{\bibinfo{volume}{100}},
  \bibinfo{pages}{026804} (\bibinfo{year}{2008}).
  
\bibitem[{\citenamefont{Romito et~al.}(2008)\citenamefont{Romito, Gefen, and
  Blanter}}]{Romito:2008}
\bibinfo{author}{\bibfnamefont{A.}~\bibnamefont{Romito}},
  \bibinfo{author}{\bibfnamefont{Y.}~\bibnamefont{Gefen}}, \bibnamefont{and}
  \bibinfo{author}{\bibfnamefont{Y.~M.} \bibnamefont{Blanter}},
  \bibinfo{journal}{Phys. Rev. Lett.} \textbf{\bibinfo{volume}{100}},
  \bibinfo{pages}{056801} (\bibinfo{year}{2008}).

\bibitem[{\citenamefont{Shpitalnik et~al.}(2008)\citenamefont{Shpitalnik,
  Gefen, and Romito}}]{Shpitalnik:2008}
\bibinfo{author}{\bibfnamefont{V.}~\bibnamefont{Shpitalnik}},
  \bibinfo{author}{\bibfnamefont{Y.}~\bibnamefont{Gefen}}, \bibnamefont{and}
  \bibinfo{author}{\bibfnamefont{A.}~\bibnamefont{Romito}},
  \bibinfo{journal}{Phys. Rev. Lett.} \textbf{\bibinfo{volume}{101}},
  \bibinfo{pages}{226802} (\bibinfo{year}{2008}).

\bibitem[{\citenamefont{Schomerus and Robinson}(2007)}]{Schomerus:2007}
\bibinfo{author}{\bibfnamefont{H.}~\bibnamefont{Schomerus}} \bibnamefont{and}
  \bibinfo{author}{\bibfnamefont{J.~P.} \bibnamefont{Robinson}},
  \bibinfo{journal}{New J. Phys.} \textbf{\bibinfo{volume}{9}},
  \bibinfo{pages}{67} (\bibinfo{year}{2007});
\bibinfo{author}{\bibfnamefont{C.}~\bibnamefont{Birchall}} \bibnamefont{and}
  \bibinfo{author}{\bibfnamefont{H.}~\bibnamefont{Schomerus}},
  \bibinfo{journal}{Phys. Rev. Lett.} \textbf{\bibinfo{volume}{105}},
  \bibinfo{pages}{026801} (\bibinfo{year}{2010}).

\bibitem[{foo({\natexlab{a}})}]{footnote:general-magnetic}
\bibinfo{note}{Generalizations into setups which include both a magnetic field
  and spin--orbit coupling can be made along the lines of Ref.
  \cite{Whitney:2008}, as $q$ only affects the accumulated dynamic phase along
  the interferometer arms.}

\bibitem[{\citenamefont{Meijer et~al.}(2002)\citenamefont{Meijer, Morpurgo, and
  Klapwijk}}]{Meijer:2002}
\bibinfo{author}{\bibfnamefont{F.~E.} \bibnamefont{Meijer}},
  \bibinfo{author}{\bibfnamefont{A.~F.} \bibnamefont{Morpurgo}},
  \bibnamefont{and} \bibinfo{author}{\bibfnamefont{T.~M.}
  \bibnamefont{Klapwijk}}, \bibinfo{journal}{Phys. Rev. B}
  \textbf{\bibinfo{volume}{66}}, \bibinfo{pages}{033107}
  (\bibinfo{year}{2002}).

\bibitem[{\citenamefont{Whitney et~al.}(2008)\citenamefont{Whitney, Shnirman,
  and Gefen}}]{Whitney:2008}
\bibinfo{author}{\bibfnamefont{R.~S.} \bibnamefont{Whitney}},
  \bibinfo{author}{\bibfnamefont{A.}~\bibnamefont{Shnirman}}, \bibnamefont{and}
  \bibinfo{author}{\bibfnamefont{Y.}~\bibnamefont{Gefen}},
  \bibinfo{journal}{Phys. Rev. Lett.} \textbf{\bibinfo{volume}{100}},
  \bibinfo{pages}{126806} (\bibinfo{year}{2008}).

\bibitem[{\citenamefont{Landauer}(1957)}]{Landauer:1957}
\bibinfo{author}{\bibfnamefont{R.}~\bibnamefont{Landauer}},
  \bibinfo{journal}{IBM J. Res. Fev.} \textbf{\bibinfo{volume}{1}},
  \bibinfo{pages}{223} (\bibinfo{year}{1957});
\bibinfo{author}{\bibfnamefont{R.}~\bibnamefont{Landauer}},
  \bibinfo{journal}{Phil. Mag.} \textbf{\bibinfo{volume}{21}},
  \bibinfo{pages}{863} (\bibinfo{year}{1970});
\bibinfo{author}{\bibfnamefont{Y.}~\bibnamefont{Imry}},
  \emph{\bibinfo{title}{Directions in Condensed Matter Physics}}
  (\bibinfo{publisher}{World Scientific, Singapore}, \bibinfo{year}{1986}).

\bibitem[{sup()}]{supplementary}
\bibinfo{note}{See Supplemental Material for (i) semiclassical analysis of the spin-orbit effect generated by a charge next to an electron wire and (ii) finite energy and magnetic fluctuations analysis..}

\bibitem[{foo({\natexlab{b}})}]{footnote:coupled}
\bibinfo{note}{In the general case we will obtain similar amplification as in
  this simplified case (see Fig.~2(b)).}

\bibitem[{\citenamefont{Ando et~al.}(1982)\citenamefont{Ando, Fowler, and
  Stern}}]{ando:1982}
\bibinfo{author}{\bibfnamefont{T.}~\bibnamefont{Ando}},
  \bibinfo{author}{\bibfnamefont{A.~B.} \bibnamefont{Fowler}},
  \bibnamefont{and} \bibinfo{author}{\bibfnamefont{F.}~\bibnamefont{Stern}},
  \bibinfo{journal}{Rev. Mod. Phys.} \textbf{\bibinfo{volume}{54}},
  \bibinfo{pages}{437} (\bibinfo{year}{1982}).

\end{thebibliography}

\appendix
\section{Supplemental Material I: Spin-orbit effect generated by a charge placed next to an electron wire}
\label{appendix1}
\begin{figure}
\includegraphics[width=7.5cm]{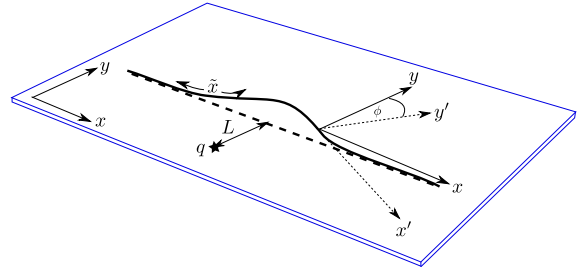}
\caption[]{\label{wire_model}
(Color online) The effect of a classical charge, $q$, on the electron's trajectory.}
\end{figure}

We calculate, here, the effect of a classical charge, $qe$, on the electrons in the adjacent arm $1$ of the AB-ring. We use an electrostatic model and plug realistic numbers for a GaAs system. The arm is treated as a straight wire in the vicinity of the charge (see Fig.~\ref{wire_model}), described by a parabolic confining potential at a distance, $l$, from the $\hat{x}$-axis, $V_{\text{har}}(x,y)=(1/2)m\omega_0^2 (y-l)^2$. We include only the lowest transverse mode by tuning $\omega_0$ such that $E_f \approx \hbar \omega_0$.
The charge is situated at the origin and is modeled with a two-dimensional electron gas electrostatic potential, $V_e(x,y)=(q e^2)/(\bar{k}\bar{q}_s^2 r^3)$, where $\bar{k}$ is the dielectric constant, $\bar{q}_s$ a screening parameter, and $r=\sqrt{x^2+y^2}$~\cite{ando:1982}. Taking $V=V_{\text{har}}+V_e$ and its derivatives $\vec{V}=(V_x,V_y),V_{xx},V_{xy},V_{yx},V_{yy}$, we find the new electron trajectory by imposing two constraints for each point along the trajectory: (i) the point is the local minimum in a given $\phi$ direction, i.e. $\vec{\nabla}\cdot\vec{V}^{\phi}=V_y-\tan(\phi)V_x=0$; (ii) the curvature is quadratic in the given direction, i.e. the off-diagonal terms in the rotated Hessian are zero $(V_{yy}-V_{xx})\tan(2\phi)+2V_{xy}=0$. Combining the two conditions we obtain the function
\begin{align}
F(x,y,q)=V_y-\left(\frac{\sqrt{1+T^2}-1}{T}\right)V_x\equiv 0\, ,
\label{functional}
\end{align}
where we have used the trigonometric relation $\tan(2\phi)=2\tan(\phi)/(1-\tan(\phi)^2)$ to define $T=\tan(2\phi)=V_{xy}/(V_{xx}-V_{yy})$. Solving Eq.~\ref{functional} perturbatively in $q$, i.e. $y\approx l+q y_1(x)$, we obtain the added trajectory due to the charge
\begin{align}
y_1(x)=-\frac{F_q|_{q=0}}{F_y|_{q=0}}\, ,
\label{trajectory}
\end{align}
where we have defined the function's derivatives $F_q,F_y$.
An electron traveling in this new trajectory will feel a position dependent minimum potential, $\Delta V({x})=V(x,l+qy_1(x))$, and a changed local confining potential in the transverse direction $m\omega_0^\prime({x})=m\omega_0+q\Delta \omega({x})=V^{\phi}_{yy}=\left((1/(1+T^2))(V_{yy}+T^2V_{xx}-2TV_{xy})\right)|_{y=l+qy_1(x)}$. This implies a local linear correction to the eigenmomenta $p_\sigma({x})=  \pm\sqrt{ \tilde{p}_0^2({x}) - g\mu_B mB_z \sigma  }$, with $\tilde{p}_0({x})=[2m((E_{\rm F}-E_0)-q(\Delta V({x})+(\hbar\Delta\omega({x}))/2))]^{1/2}$.
The overall additional phase accumulated by the electron in the wire due to the presence of the nearby charge is obtained by integrating over the difference in acquired dynamical phases over the new and old trajectory,
\begin{align}
\int_\gamma \frac{p_\sigma(\gamma)\cdot d\vec{l}}{\hbar}&=\int_{-\infty}^{\infty}\frac{dx}{\hbar} (p_\sigma({x})\sqrt{1+q^2 (y_1^\prime(x))^2}-p_\sigma)\nonumber\\
&  \approx -qe  p_{\bar{\sigma}} \eta\, .
\end{align}
Here
\begin{align}
\eta\equiv \frac{m}{e\hbar\mathcal{P}^2}\int_{-\infty}^{\infty}dx \Delta V({x})+\frac{\hbar\Delta\omega({x})}{2}\, ,
\end{align}
takes into account the detailed form of the potentials, and we have used the fact the change in the length of the modified trajectory is negligible ($\sim q^2$).
Additionally, the new trajectory encloses a different area, $\Delta{\cal A}=\int_{-\infty}^{\infty}dx\; y_1(x)$, which will affect the accumulated AB-phase $\Phi_{\text{AB}}\rightarrow \Phi_{\text{AB}}+\delta e$.

In GaAs $E_f \cong 10$meV, therefore, we choose $\omega_0 \sim 15$THz, which is equivalent to a wire width of $\sim 19.5$nm. The dielectric constant is $\bar{k}=12.9$ and the screening length $l_s=1/q_s\sim 8.5$nm. We take the charge's distance from the wire to be $l=3l_s\sim 25.5$nm with a wire length $L=1\mu$m. For a magnetic field of $B_z \sim 1$T, this results in $(\bar{p}/\hbar)\sim 7\cdot 10^{-2}$(1/nm), $(\Delta p)/\hbar\sim 1\cdot 10^{-4}$(1/nm), $\hbar e\eta\sim 20.7$nm,  $\delta e\sim 1.5\cdot 10^{-2}$.

\section{Supplemental Material II: Finite energy and magnetic fluctuations}
The WV signal is sensitive to temperature, voltage bias, and magnetic field fluctuations. However, it still survives to generate an amplification in reasonable ranges of parameters. We consider the effect of each of these factors separately. The energy and temperature dependence of the transmission is calculated by plugging, $t_{L\rightarrow R}\rightarrow t_{L\rightarrow R}(E)$, into Eq.~\eqref{landauer} and performing the integrals. Similarly, with $t_{L\rightarrow R}\rightarrow t_{L\rightarrow R}(B)$, we integrate over a Gaussian distribution of width $\Delta B$ around the original magnetic field $B_0$.
To solve the energy (magnetic field) integral, we linearize the momenta around $E_f$ ($B_0$), i.e. $\bar{p}\rightarrow \bar{p}(E)=\bar{p}(1+(m/\mathcal{P}^2)E)$, $\Delta p\rightarrow \Delta p(E)=\Delta p(1-(m/\mathcal{P}^2)E)$ ($\bar{p}\rightarrow \bar{p}(B)=\bar{p}-\Delta p(g\mu_B m/2\mathcal{P}^2)B$, $\Delta p\rightarrow \Delta p(B)=\Delta p+\bar{p}(g\mu_B m/2\mathcal{P}^2)B$
).
Additionally, $\Phi_{\text{AB}},\delta\rightarrow \Phi_{\text{AB}}(B),\delta(B)$ depends on the magnetic field fluctuations.
In Eq.~\eqref{trasmissione} the dependence of $\Delta T_q$ on energy and magnetic field is due to the following terms:
$|\langle \mathbf{n}_R | \mathbf{n}_L' \rangle |^2=\sum_\sigma|\langle \sigma | \mathbf{n}_R \rangle |^2|\langle \sigma | \mathbf{n}_L \rangle |^2+2\re  \left\{e^{i 2\Delta p(E,B) L} \langle \mathbf{n}_L|\dn \rangle\langle \dn | \mathbf{n}_R \rangle \langle \mathbf{n}_R\up \rangle\langle \up | \mathbf{n}_L \rangle \right\}$, $|\langle \phi_f |\phi_i \rangle|^2 {}_f \langle \hat{A}\rangle_i =t_1^2+t_1 t_2e^{-i\tilde{\Phi}(E,B)}$, and $|\langle \mathbf{n}_R | \mathbf{n}_L' \rangle |^2{}_R\langle \hat{\sigma}_z\rangle_L=\sum_\sigma \sigma|\langle \sigma | \mathbf{n}_R \rangle |^2|\langle \sigma | \mathbf{n}_L \rangle |^2+2i\im  \left\{e^{i 2\Delta p(E,B) L} \langle \mathbf{n}_L|\dn \rangle\langle \dn | \mathbf{n}_R \rangle \langle \mathbf{n}_R\up \rangle\langle \up | \mathbf{n}_L \rangle \right\}$. Performing the integral over the $E$ and $B$ windows, we obtain
\begin{widetext}
\begin{eqnarray}
    \label{trasmissionen}
&&    \Delta I_q(T,V) = -2 q eV(\delta+\eta\bar{p})t_1 t_2\Bigg(\sum_\sigma|\langle \sigma | \mathbf{n}_R \rangle |^2|\langle \sigma | \mathbf{n}_L \rangle |^2 \im \left\{e^{-i\tilde{\Phi}}K\left(T,V,\frac{\bar{p}\Delta L}{\hbar}\right)\right\}
+|\langle \mathbf{n}_L|\dn \rangle\langle \dn | \mathbf{n}_R \rangle \langle \mathbf{n}_R\up \rangle\langle \up | \mathbf{n}_L \rangle|\times\nonumber\\
&&\Big(\im \left\{e^{i( 2\frac{\Delta p L}{\hbar}-\varphi-\tilde{\Phi})}K\left(T,V,-\frac{\Delta p L+\bar{p}\Delta L}{\hbar}\right)\right\}-\im\left\{e^{i( 2\frac{\Delta p L}{\hbar}-\varphi+\tilde{\Phi})}K\left(T,V,\frac{\bar{p}\Delta L-\Delta p L}{hbar}\right)\right\}\Big)\Bigg)\nonumber\\
&&+2q eV\eta\Delta p\Bigg( 2t_1^2|\langle \mathbf{n}_L|\dn \rangle\langle \dn | \mathbf{n}_R \rangle \langle \mathbf{n}_R\up \rangle\langle \up | \mathbf{n}_L \rangle|\im \left\{e^{i( 2\frac{\Delta p L}{\hbar}-\varphi)}K\left(T,V,-\frac{\Delta p L}{\hbar}\right)\right\}\nonumber\\
&&+t_1 t_2\Bigg[\sum_\sigma \sigma|\langle \sigma | \mathbf{n}_R \rangle |^2|\langle \sigma | \mathbf{n}_L \rangle |^2\im \left\{e^{-i\tilde{\Phi}}K\left(T,V,\frac{\bar{p}\Delta L}{\hbar}\right)\right\}
+|\langle \mathbf{n}_L|\dn \rangle\langle \dn | \mathbf{n}_R \rangle \langle \mathbf{n}_R\up \rangle\langle \up | \mathbf{n}_L \rangle|\times\nonumber\\
&&\left(\im \left\{e^{i( 2\frac{\Delta p L}{\hbar}-\varphi-\tilde{\Phi})}K\left(T,V,-\frac{\Delta p L+\bar{p}\Delta L}{\hbar}\right)\right\}+\im \left\{e^{i( 2\frac{\Delta p L}{\hbar}-\varphi+\tilde{\Phi})}K\left(T,V,\frac{\bar{p}\Delta L-\Delta p L}{\hbar}\right)\right\}\right)\Bigg]\Bigg)\, ,\\
&&    \Delta I_q(\Delta B) = -2 q eV(\delta+\eta\bar{p})t_1 t_2\Bigg(\sum_\sigma|\langle \sigma | \mathbf{n}_R \rangle |^2|\langle \sigma | \mathbf{n}_L \rangle |^2 \im \left\{e^{-i\tilde{\Phi}}F\left(\Delta B,e{\cal A}/(g \mu_{\rm B}\hbar)-\frac{m \Delta p\Delta L}{2\hbar\mathcal{P}^2}\right)\right\}\nonumber\\
&&+|\langle \mathbf{n}_L|\dn \rangle\langle \dn | \mathbf{n}_R \rangle \langle \mathbf{n}_R\up \rangle\langle \up | \mathbf{n}_L \rangle|\Big(\im \left\{e^{i( 2\frac{\Delta p L}{\hbar}-\varphi-\tilde{\Phi})}F\left(\Delta B,e{\cal A}/(g \mu_{\rm B}\hbar)+\frac{m\left(2\bar{p} L-\Delta p\Delta L\right)}{2\hbar\mathcal{P}^2}\right)\right\}\nonumber\\
&&-\im \left\{e^{i( 2\frac{\Delta p L}{\hbar}-\varphi+\tilde{\Phi})}F\left(\Delta B,-e{\cal A}/(g \mu_{\rm B}\hbar)+\frac{m\left(2\bar{p} L+\Delta p\Delta L\right)}{2\hbar\mathcal{P}^2}\right)\right\}\Big)\Bigg)\nonumber\\
&&+2q eV\eta\Delta p\Bigg( 2t_1^2|\langle \mathbf{n}_L|\dn \rangle\langle \dn | \mathbf{n}_R \rangle \langle \mathbf{n}_R\up \rangle\langle \up | \mathbf{n}_L \rangle|\im \left\{e^{i( 2\frac{\Delta p L}{\hbar}-\varphi)}F\left(\Delta B,\frac{m \bar{p} L}{2\hbar\mathcal{P}^2}\right)\right\}\nonumber\\
&&+t_1 t_2\Bigg[\sum_\sigma \sigma|\langle \sigma | \mathbf{n}_R \rangle |^2|\langle \sigma | \mathbf{n}_L \rangle |^2\im \left\{e^{-i\tilde{\Phi}}F\left(\Delta B,e{\cal A}/(g \mu_{\rm B}\hbar)-\frac{m \Delta p\Delta L}{2\hbar\mathcal{P}^2}\right)\right\}
+|\langle \mathbf{n}_L|\dn \rangle\langle \dn | \mathbf{n}_R \rangle \langle \mathbf{n}_R\up \rangle\langle \up | \mathbf{n}_L \rangle|\times\nonumber\\
&&\Big(\im \left\{e^{i( 2\frac{\Delta p L}{\hbar}-\varphi-\tilde{\Phi})}F\left(\Delta B,e{\cal A}/(g \mu_{\rm B}\hbar)+\frac{m\left(2\bar{p} L-\Delta p\Delta L\right)}{2\hbar\mathcal{P}^2}\right)\right\}\nonumber\\
&&+\im \left\{e^{i( 2\frac{\Delta p L}{\hbar}-\varphi+\tilde{\Phi})}F\left(\Delta B,-e{\cal A}/(g \mu_{\rm B}\hbar)+\frac{m\left(2\bar{p} L+\Delta p\Delta L\right)}{2\hbar\mathcal{P}^2}\right)\right\}\Big)\Bigg]\Bigg)\, ,
\label{trasmissioneb}
    \end{eqnarray}
\end{widetext}
where we have used the azimuth phase difference between the pre- and post-selected spin orientations $\varphi$, the function $K(T,V,N)=\frac{2\pi }{\beta eV}e^{i\frac{eV m N}{2\mathcal{P}^2}} \frac{\sin\left(\frac{ eV m N}{2\mathcal{P}^2}\right)}{\sinh\left(\frac{\pi m N}{\mathcal{P}^2\beta}\right)}$, and $F(\Delta B,N)=e^{-\frac{1}{2} \left(\Delta B N\right)^2}$.

Similar temperature, voltage bias, and magnetic field fluctuations dependence expressions can be written for the {}``noise'' of our device, $\Delta T_\xi$ (cf. textual equation above Eq.~\eqref{amplificazione}). The fact that the {}``signal'', $\Delta T_q$ (Eq.~\eqref{trasmissione}), and the {}``noise'' are both attenuated by finite temperature and/or voltage can be traced back to the following fact. Consider first the {}``signal''. It consists of a sum of products of terms which vary sensitively with $\bar{p}$ and $\Delta p$ (they both appear in phase factors such as $L\cdot \Delta p$, $\Delta L \cdot \bar{p}$, and combinations thereof). Averaging over different energies (as implied by finite $T,V$) means that we average over those phase factors, which leads to attenuation. The combination of the $\Delta p, \bar{p}$ phase sensitive factors is different between the {}``signal'' and the {}``noise'' terms, hence the different attenuation and the smearing of the amplification (see Figs.~\ref{Fig:3},\ref{Fig:4}).

\end{document}